\date{\today}

\documentclass[aps,prl,twocolumn,showpacs,superscriptaddress,groupedaddress]{revtex4}  
\usepackage{dcolumn,graphicx,amsmath,amssymb,algorithm,algpseudocode}
\usepackage{todonotes}
\usepackage{qcircuit}

\begin{document}

\title{
Quantum Computation of Electronic Transitions using a Variational Quantum
Eigensolver
}

\author{Robert M. Parrish}
\email{rob.parrish@qcware.com}
\affiliation{
Department of Chemistry and the PULSE Institute, Stanford University, Stanford, CA 94305
    }
\affiliation{
SLAC National Accelerator Laboratory, Menlo Park, CA 94025
}
\affiliation{
QC Ware Corporation, Palo Alto, CA 94301
}

\author{Edward G. Hohenstein}
\affiliation{
Department of Chemistry and the PULSE Institute, Stanford University, Stanford, CA 94305
    }
\affiliation{
SLAC National Accelerator Laboratory, Menlo Park, CA 94025
}

\author{Peter L. McMahon}
\affiliation{
E.\,L.~Ginzton Laboratory, Stanford University, Stanford, CA 94305
    }
\affiliation{
QC Ware Corporation, Palo Alto, CA 94301
}

\author{Todd J. Mart\'inez}
\affiliation{
Department of Chemistry and the PULSE Institute, Stanford University, Stanford, CA 94305
    }
\affiliation{
SLAC National Accelerator Laboratory, Menlo Park, CA 94025
}
\email{Todd.Martinez@gmail.com}

\begin{abstract} 
We develop an extension of the variational quantum eigensolver (VQE) algorithm
-- multistate, contracted VQE (MC-VQE) -- that allows for the efficient
computation of the transition energies between the ground state and several
low-lying excited states of a molecule, as well as the oscillator strengths
associated with these transitions. We numerically simulate MC-VQE by computing
the absorption spectrum of an \emph{ab initio} exciton model of an
18-chromophore light-harvesting complex from purple photosynthetic bacteria. 
\end{abstract}

\pacs{03.67.Ac,31.10.+z,31.15.-p}

\maketitle

The accurate modeling of the many-body interactions in the ground and
excited-state solutions of the electronic Schr\"odinger equation is a
prerequisite for the quantitative prediction of molecular physical phenomena such
as light harvesting. Using classical computers, this problem scales formally as
the factorial of the number of involved electrons \cite{PurpleBook}, via the
solution of the full configuration interaction (FCI) equations, though many
polynomial-scaling approximations such as density functional theory \cite{
Hohenberg:1964:B864,
Kohn:1965:A1133,
Runge:1984:997,
Koch:2001:DFT}
(DFT), coupled cluster theory \cite{
Cizek:1966:4256,
Purvis:1982,
Crawford:2007:33,
Bartlett}
(CC), density matrix renormalization group \cite{
White:1992:2863,
Chan:2011:465}
(DMRG),
adaptive and/or stochastic configuration interation methods \cite{
bender1969studies,
huron1973iterative,
booth2009fermion,
cleland2010communications,
holmes2016heat,
Schriber:2016:161106,
schriber2017adaptive}
(CIPSI and variants),
and semistochastic coupled cluster methods \cite{
deustua2017converging,
deustua2018communication},
have been developed to combat this problem. Recently, there has been a
surge of interest in using quantum computers to naturally solve the many-body
electronic structure problem through methods such as the iterative phase
estimation algorithm \cite{
Abrams:1997:2586,
Abrams:1999:5162,
Aspuru:2005:1704,
Lanyon:2010:106,
Wecker:2014:022305,
Tubman:2018:X}
(IPEA) or the variational quantum eigensolver \cite{ 
Peruzzo:2014:4213,
McClean:2016:023023,
OMalley:2016:031007,
Kandala:2017:242,
McClean:2017:X,
Romero:2018:104008}
(VQE), However, the quartic-scaling complexity in number of molecular orbitals
of the second-quantized electronic Hamiltonian, coupled with the overhead of
encoding the fermionic antisymmetry of the electrons through the Jordan-Wigner
\cite{Jordan:1928:631,Ortiz:2001:022319} (JW), Bravyi-Kitaev
\cite{Bravyi:2002:210,Seeley:2012:224109} (KB), or superfast Bravyi-Kitaev
\cite{Setia:2017:X,Setia:2018:X} (SFKB) transformations, implies that rather
long circuit depths will be required to directly model the electronic structure
problem. We also point out a recent approach \cite{Babbush:2017:X,
Kivlichan:2018:110501,Motta:2018:X} that might formally reduce this complexity
to quadratic or linear via a tensor hypercontraction representation
\cite{Hohenstein:2012:044103,Parrish:2012:224106,Parrish:2013:132505} of the
potential. In the present work, we explore a domain- and problem-specific means
to reduce the complexity of the representation of the electronic structure
problem in quantum computing: an \emph{ab initio} exciton model \cite{
Sisto:2014:2857,
Sisto:2017:14924,
Li:2017:3493,
Morrison:2014:5366,
Morrison:2015:4390}. For large-scale photoactive complexes consisting of a
number of nonbonded chromophore units, the \emph{ab initio} exciton model
compresses the details of the electronic structure on each chromophore into a
handful of monomer electronic states. The determination of the full
configuration interaction wavefunctions describing the mixing of monomer
electronic states in the full complex remains a formidable task - here we show
that this might be a natural computational task for a near-term quantum
computer.

Another area that deserves exploration is the development of efficient quantum
algorithms for the even-handed treatment of ground- and excited-state energies
and transition properties, e.g., for the computation of absorption spectra.
There exist IPEA-type algorithms for excited states, such as the WAVES protocol
\cite{Santagati:2018:eaap9646} or the variational swap test \cite{Endo:2018:X},
but we focus on VQE-type methods here. Most existing VQE-type quantum algorithms
are ``state specific,'' meaning that they optimize the VQE parameters for one
state at a time. Examples include the folded spectrum (FS) method
\cite{Peruzzo:2014:4213}, which requires the observation of the square of the
Hamiltonian, or the orthogonality-constrained VQE (OC-VQE) method
\cite{Higgott:2018:X,Lee:2018:JCTC} which applies a penalty term to remove
contaminants from lower-lying states. Another, more-global approach is the
quantum subspace expansion (QSE-VQE)
\cite{McClean:2017:042308,Colless:2018:011021}, which first performs VQE to
determine the ground state, and then determines the excited states by classical
diagonalization in a basis of response states.  QSE-VQE treats all the excited
states on a similar footing, but by construction favors the ground state, and
requires the determination of three- and four-particle density matrices through
high-order Pauli measurements.

\textit{MC-VQE} - Inspired by the mixed quantum/classical strategy of QSE-VQE
(particularly the final classical diagonalization step), we have developed a new
multistate, contracted variant of VQE (MC-VQE), which aims to (1) treat the
ground and a handful of excited states on the same footing (2) minimize the size
of the classical subspace that must be diagonalized and (3) provide for the
straightforward computation of transition properties such as oscillator
strengths. MC-VQE takes the following ansatz for a number ($N_{\Theta}$) of
eigenstates of interest,
\begin{equation}
\label{eq:vqe}
|\Psi_{\Theta} \rangle
\equiv
\hat U 
\sum_{\Theta'}
| \Phi_{\Theta'} \rangle
V_{\Theta' \Theta}
.
\end{equation}
Here $| \Phi_{\Theta} \rangle$ are a set of contracted, orthonormal
``reference'' states, which are obtained by solving a classical electronic
structure problem such as configuration interaction singles (CIS). By
contracted, we mean that these reference states are generally taken to be a
linear combination of Hilbert-space configurations - ideally this will allow the
reference states to be reasonably accurate approximations to the exact
eigenstates. As will be seen, all that we will require is that we have an
efficient quantum circuit to prepare the ``diagonal'' state $|\Phi_{\Theta}
\rangle$ and the ``interfering'' state $(|\Phi_{\Theta} \rangle \pm
|\Phi_{\Theta'} \rangle) / \sqrt{2}$. For CIS reference states, this is possible
- see the Supplemental Material for a detailed circuit \cite{SuppNote} which
  generalizes a previously known circuit for $|W_N\rangle$ states
\cite{Diker:2016:W}.

The operator $\hat U (\{ \eta \})$ is the VQE entangler matrix, an orthogonal
Hilbert-space matrix constructed from a set of two-qubit entangling operators
whose set of parameters $\{ \eta \}$ will be chosen to maximally decouple $\{ |
\Phi_{\Theta'} \rangle \}$ from the rest of the Hilbert space, i.e., to
approximately block diagonalize the Hamiltonian.  The matrix $V_{\Theta'
\Theta}$ is an $N_{\Theta} \times N_{\Theta}$ orthogonal matrix that describes
the rotation of the entangled contracted states $ \{ | \chi_{\Theta'} \rangle
\equiv \hat U |\Phi_{\Theta'} \rangle \} $ to the approximate eigenbasis $\{ |
\Psi_{\Theta} \rangle \}$. This matrix can be determined by classical
diagonalization of the entangled contracted Hamiltonian,
\begin{equation}
H_{\Theta'' \Theta'}
V_{\Theta' \Theta}
=
V_{\Theta'' \Theta}
E_{\Theta}
\ : \
V_{\Theta' \Theta}
V_{\Theta' \Theta''}
=
\delta_{\Theta \Theta''}
.
\end{equation}
The eigenvalues $E_{\Theta}$ are the Ritz approximations to the exact
eigenvalues. The entangled contracted Hamiltonian is,
\begin{equation}
H_{\Theta \Theta'}
\equiv
\langle \Phi_{\Theta} | \hat U^{\dagger} \hat H \hat U | \Phi_{\Theta'} \rangle
.
\end{equation}
The diagonal matrix elements can be evaluated by partial tomography measurements
in a quantum computer, as is done in standard VQE:
\begin{equation}
\label{eq:Heff}
H_{\Theta \Theta}
=
\langle \Phi_{\Theta} | \hat U^{\dagger} \hat H \hat U | \Phi_{\Theta} \rangle
.
\end{equation}
The (real) off-diagonal matrix elements can also be obtained from observable
quantities:
\[
2 H_{\Theta \neq \Theta'}
=
\left (
\langle \Phi_{\Theta} |
+
\langle \Phi_{\Theta'} |
\right )
\hat U^{\dagger}
\hat H
\hat U
\left (
| \Phi_{\Theta} \rangle
+
| \Phi_{\Theta'} \rangle
\right ) / 2
\]
\begin{equation}
\label{eq:Htrans}
-
\left (
\langle \Phi_{\Theta} |
-
\langle \Phi_{\Theta'} |
\right )
\hat U^{\dagger}
\hat H
\hat U
\left (
| \Phi_{\Theta} \rangle
-
| \Phi_{\Theta'} \rangle
\right ) / 2
.
\end{equation}
This highlights the need for quantum circuits to prepare the ``interfering''
state $(|\Phi_{\Theta} \rangle \pm |\Phi_{\Theta'} \rangle) / \sqrt{2}$. 

The parameters of the MC-VQE entanglement circuit should be chosen to maximally
decouple the full set of approximate eigenstates $\{ | \Psi_{\Theta} \rangle
\}$ from the rest of the Hilbert space. This can be accomplished in a 2-norm
sense in the Hamiltonian by optimizing the parameters of the VQE entangler
operator to minimize the state-averaged energy,
\begin{equation}
\bar{E}
=
\frac{1}{N_{\Theta}}
\sum_{\Theta}
E_{\Theta}
=
\frac{1}{N_{\Theta}}
\sum_{\Theta}
H_{\Theta \Theta}
.
\end{equation}
The second equality follows from the definition of the trace, and shows that the
minimization of the state-averaged energy is equivalent to the minimization of
the sum of diagonal contracted Hamiltonian matrix elements.

\begin{figure}[h!]
\begin{center}
\includegraphics[width=3.2in]{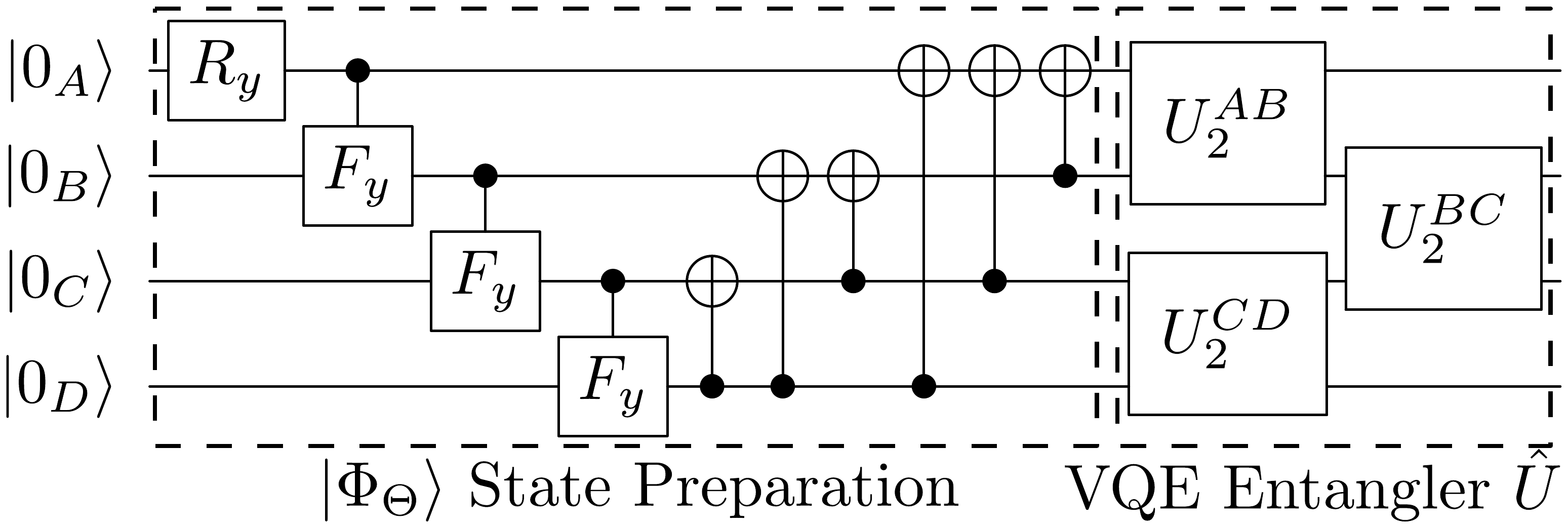}
\caption{Example MC-VQE quantum circuit for $N=4$ linear exciton model. The
first stage prepares contracted CIS reference states $|\Phi_{\Theta} \rangle$
[or interference variations $(|\Phi_{\Theta} \rangle \pm | \Phi_{\Theta'}
\rangle) / \sqrt{2}$ thereof] specified by rotation angles in the $R_y$ and
$F_y$ gates. The second stage applies the many-body VQE entangler $\hat U$
specified through a polynomial number of rotation angles 2-body $U_2$ entangler
gates. 1- and 2-body Pauli measurements of this circuit then determine the
entangled contracted Hamiltonian matrix elements $H_{\Theta \Theta'}$.}
\label{fig:vqe}
\end{center}
\end{figure}

Overall the MC-VQE algorithm has four stages:
\begin{enumerate}
\item Classically solve CIS or some other polynomial-scaling
electronic structure problem to ``sketch out'' the shapes of the relevant states by
determining the contracted reference states $\{ | \Phi_{\Theta} \rangle \}$.
\item Vary the parameters of the VQE entangler operator to optimize the
state-averaged energy $\bar E = (1/N_{\Theta}) \sum_{\Theta} H_{\Theta \Theta}$.
\item For the converged VQE entangler operator, observe the reference-state
Hamiltonian $H_{\Theta \Theta'}$ using sums and differences of Hamiltonian
expectations of interference states.
\item Classically diagonalize $H_{\Theta \Theta'}$ to obtain the Ritz estimates
of the eigenstates and eigenvalues.
\end{enumerate}
A schematic of the quantum circuit needed to prepare a CIS state
$|\Phi_{\Theta}\rangle$ and apply the VQE entangler $\hat U$ is shown in Figure
\ref{fig:vqe} - details of this circuit are available in the Supplemental
Material \cite{SuppNote}. Overall, the MC-VQE approach has a number of unique
features relative to established excited-state VQE approaches such as quantum
subspace expansion (QSE-VQE) \cite{McClean:2017:042308}:
\begin{itemize}
\item The VQE entangler $\hat U$ is optimized in a state-averaged manner,
providing a balanced treatment of ground and excited states, i.e., all states
are computed to approximately equal accuracy.
\item The optimization of the VQE entangler $\hat U$ requires only the
measurement of $N_{\Theta}$
diagonal matrix elements $H_{\Theta \Theta}$. The determination of the
$N_{\Theta}^2$ off-diagonal matrix elements $H_{\Theta \neq \Theta'}$ can be
done separately, after the VQE entangler parameters have been optimized.
\item Higher-order density matrices are not required. 
\end{itemize}

Note that the eigenstates can be reexpressed as
$
| \Psi_{\Theta} \rangle
\equiv
\hat U
| \Gamma_{\Theta} \rangle
$
where $\{ |\Gamma_{\Theta} \rangle \equiv \sum_{\Theta'} | \Phi_{\Theta'}
\rangle V_{\Theta' \Theta}  \}$ are rotated reference states.  The algorithm
above is quite general - we present a demonstration below for the case of the
\emph{ab initio} exciton model, but it is clear that this approach might be
immediately applicable to the efficient computation of excited states in
fermionic electronic structure computations.  Transition properties (such as the
transition dipole moment, needed for computing the absorption spectrum) can also
be computed by substituting the desired operator $\hat O$ in place of $\hat H$
in Equation \ref{eq:Htrans}.

It is worth noting that MC-VQE can be roughly pictured either as a generator
of the wavefunction ansatz of Equation \ref{eq:vqe} or as a means to observe
the elements of the unitarily-transformed effective Hamiltonian of Equation
\ref{eq:Heff}, wherein the VQE entangler operator $\hat U$ acts as a
wave operator \cite{durand1983direct,maynau1983direct}. 

\textit{Ab Initio Exciton Model} - Consider a set of $N$ chromophoric monomers,
each
labeled by index $A$, which are arranged in a particular nuclear geometry in a
photoactive complex. In isolation, the chromophores are usually characterized by a
constant number of photoactive electronic states, regardless of the number of
electrons in the monomer (often between two and four states are photoactive in
the visible spectrum in the monomer: the ground and the first few singlet excited
states). If the monomers are sufficiently far apart in the full photoactive
complex (e.g., if they are at noncovalent separations due to embedding in a
protein scaffold), the strict considerations of fermionic antisymmetry can be
relaxed without loss of accuracy, and the full complex electronic eigenstates
can be computed as a configuration interaction of direct products of monomer
states. I.e., for electronic state $\Theta$ in a system where each chomophoric
monomer is characterized by the ground state $|0_A\rangle$ and the first excited
state $|1_A\rangle$ (a restriction we make from here onward to facilitate ease
of mapping to qubits), the electronic states are,
\begin{equation}
| \Psi_{\Theta} \rangle
=
\sum_{p_{0}, q_{1} \ldots \in [0,1]}
C_{p_{0} q_{1} \ldots r_{N-1}}^{\Theta}
|p_{0}\rangle
\otimes
|q_{1}\rangle
\otimes
\ldots
\otimes
|r_{N-1}\rangle
.
\end{equation}
Typically, we wish to find these adiabatic electronic states, e.g., to determine
the energy gaps and oscillator strengths in the system as a proxy for the
electronic absorption spectrum. Formally, this requires diagonalization of the
exciton Hamiltonian, which can straightforwardly be written in Pauli matrix
notation for the special case considered here of a photoactive system with two
electronic states per monomer,
\begin{equation}
\label{eq:Hpauli}
\hat H
=
\mathcal{E}
+
\mathcal{H}^{(1)}
+
\mathcal{H}^{(2)}
=
\mathcal{E} \hat I 
+ 
\sum_{A}
\mathcal{Z}_{A}
\hat Z_{A}
+ 
\mathcal{X}_{A}
\hat X_{A}
\end{equation}
\[
+
\sum_{A > B}
\mathcal{XX}_{AB}
\hat X_{A} \otimes \hat X_{B}
+
\mathcal{XZ}_{AB}
\hat X_{A} \otimes \hat Z_{B}
\]
\[
+
\mathcal{ZX}_{AB}
\hat Z_{A} \otimes \hat X_{B}
+
\mathcal{ZZ}_{AB}
\hat Z_{A} \otimes \hat Z_{B}
.
\]
The choice of Hamiltonian matrix elements $\{ \mathcal{Z}_{A}, \mathcal{X}_{A},
\mathcal{ZZ}_{AB}, \mathcal{ZX}_{AB}, \mathcal{XZ}_{AB}, \mathcal{XX}_{AB} \}$
for a given photoactive complex is an interesting art. Choosing these parameters
empirically to match experiment or other reference data is the crux of the
phenomonological Frenkel-Davydov exciton model
\cite{Frenkel:1931:17,Davydov:1964:145}. Recently, we introduced a new \emph{ab
initio} exciton model approach 
\cite{
Sisto:2014:2857,
Sisto:2017:14924,
Li:2017:3493,
Morrison:2014:5366,
Morrison:2015:4390}, 
in which the parameters of the exciton model are determined explicitly by
high-level \emph{ab initio} computations on the isolated monomers, under the
assumption of sufficient monomer separations to relax the fermionic antisymmetry
constraint. We have extended the \emph{ab initio} exciton model to treat full
non-adiabatic dynamics through the development of analytical gradients/coupling
vectors \cite{Li:2017:3493,Sisto:2017:14924} and have increased the basis set to
include both local and charge-transfer excitations \cite{Li:2017:3493}.  

In this \emph{ab initio} exciton model the Hamiltonian matrix elements in
Equation \ref{eq:Hpauli} all have distinct physical origins: $\mathcal{E}$ is
the mean-field energy, $\mathcal{Z}_{A}$ is roughly (half) of the difference
between the ground and excited state energy of monomer $A$, $\mathcal{XX}_{AB}$
is the transition-dipole--transition-dipole interaction and $\mathcal{ZZ}_{AB}$
is the difference-dipole--difference-dipole interaction between monomers $A$ and
$B$, and $\mathcal{XZ}_{AB}$ and $\mathcal{ZX}_{AB}$ are
transition-dipole--difference-dipole interaction cross terms. $\mathcal{Z}_{A}$
and $\mathcal{X}_{A}$ carry Fock-matrix like dressings from the mean-field
electrostatic environment of the system. A full definition of the matrix
elements is available in the Supplemental Material \cite{SuppNote}.

Diagonalizing this Hamiltonian to obtain the eigenstates $\{ | \Psi_{\Theta}
\rangle \}$, even for a model of this simplicity, is difficult
classically due to the $2^N$ dimension of the Hilbert space $
|p_{0}\rangle
\otimes
|q_{1}\rangle
\otimes
\ldots
\otimes
|r_{N-1}\rangle$.
To highlight this, we point out that this part of the problem is usually solved
classically in a highly restricted Hilbert space where only single excitations
are allowed \cite{
Sisto:2014:2857,
Sisto:2017:14924,
Li:2017:3493}:
for many energy-transfer applications this may be reasonable, but will be
incapable of describing the conical intersection between the ground and
lowest-excited states \cite{Levine:2006:1039}. However, it is apparent that the
\emph{ab initio} exciton Hamiltonian is entirely isomorphic to an extended
spin-lattice Hamiltonian. Therefore, existing technologies for the quantum
simulation of spin-lattice Hamiltonians should provide utility for this problem.
Below, we demonstrate the potential for this mapping by simulating the quantum
computation of the absorption spectrum of a large photoactive complex using
MC-VQE. Note that we are not the first to propose a crossover between exciton
models for photoactive complexes and spin-lattice models in qubits: there have
been myriad prior studies using phenomenological exciton models to theoretically
characterize \cite{
Plenio:2008:113019,
Caruso:2009:09B612,
Caruso:2010:062346}
or physically simulate \cite{
Mostame:2012:105013,
Mostame:2017:44,
Potovcknik:2018:904,
Wang:2018}
the exciton energy transfer (EET) process in open systems such as the
Fenna-Matthews-Olsen (FMO) complex. However, the emphasis in the prior
literature has been on the modeling of the disappative non-adiabatic dynamics of
EET through coupling with the protein/solvent environment in an effective way
(via effective phonon coupling approaches such as the Holstein model). In our
approach, we emphasize the accurate \emph{ab initio} computation of the
electronic absorption spectrum at a given nuclear configuration, as a
prerequisite for direct non-adiabatic dynamics simulations.

\textit{Demonstration} - MC-VQE circuits were implemented in our in-house quantum
simulator package, \textsc{Quasar}. All aspects of state preparation, VQE
entanglement, and casting of transition matrix elements as difference
observables were performed in the simulator, though 1- and 2-body Pauli
expectation values were evaluated through contractions of wavefunction
amplitudes (equivalent to infinite averaging of discrete Pauli measurements),
and noise/error channels were not modeled. CIS is solved classically in the
basis of the reference and all singly-excited configurations.  We avoid the
``barren plateaus'' issue of locating optimized VQE parameters
\cite{McClean:2018:X} by finding a tightly converged and near-global-optimal
solution for the 108 MC-VQE parameters which is directly downhill from a
zero-entanglement guess in 14 L-BFGS iterations, using finite-difference
gradients \cite{SuppNote}.

\begin{figure}
\begin{center}
\includegraphics[width=3.2in]{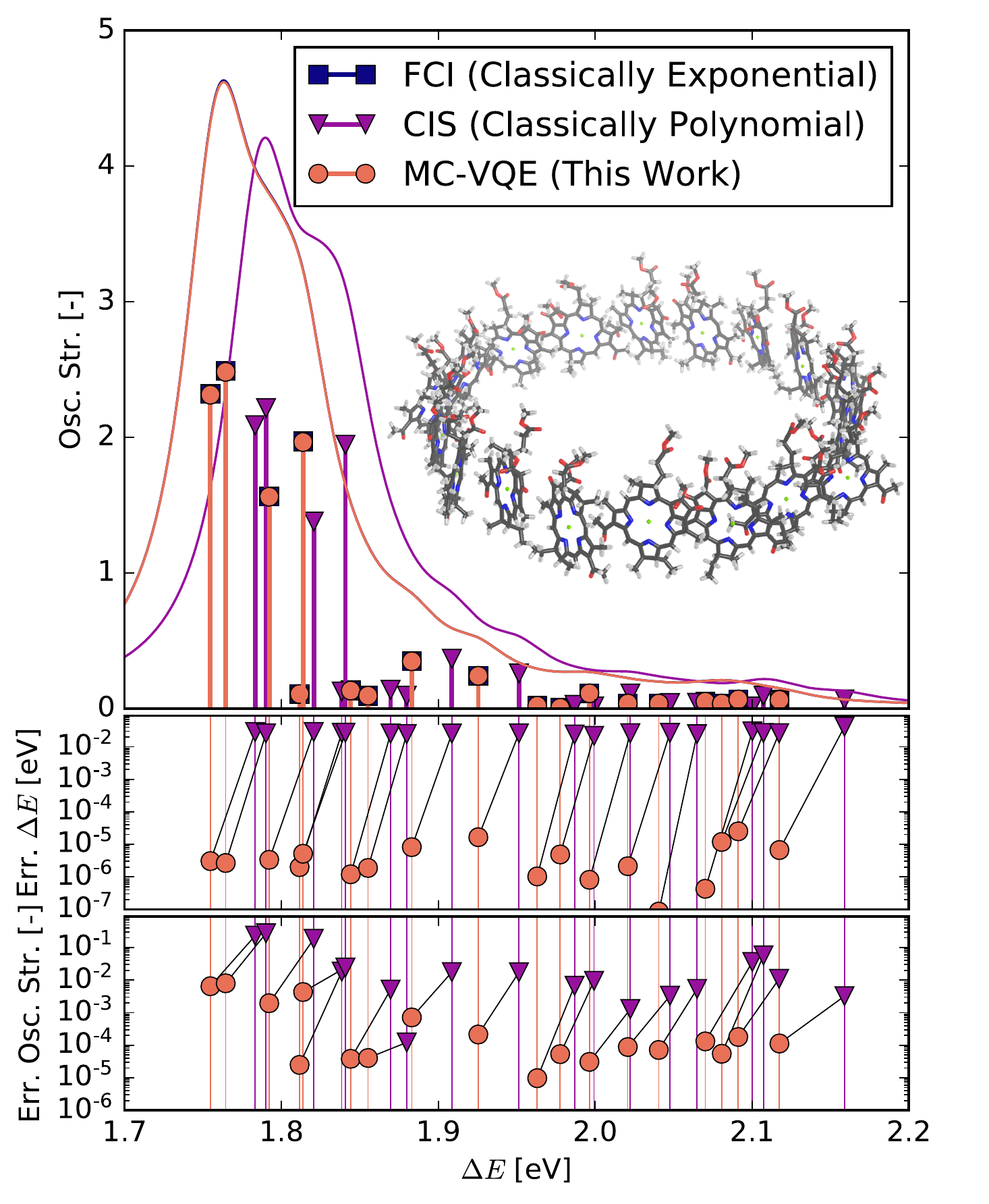}
\caption{(color online). Top - Simulated absorption spectrum of $N=18$ cyclical
LH2 B850 ring complex (geometry depicted in inset), computed from the excitation
energies and oscillator strengths of the lowest 18 electronic transitions,
depicted as vertical sticks.  The envelope of the absorption spectrum is
sketched by broadening the contribution from each transition with a Lorentzian
with width of $\delta=0.05$ eV. The simulated MC-VQE and reference FCI results
are visually indistinguishable. Middle
- errors in excitation energies. Bottom - errors in oscillator strengths. Middle
  and bottom - thin lines are a guide for the eye.
}
\label{fig:abs}
\end{center}
\end{figure}

For a practical test case, an \emph{ab initio} exciton
model was constructed for the $N=18$ cyclical LH2
B850 ring complex of the purple photosynthetic bacteria - the specific geometry
is provided in the Supplemental Material \cite{SuppNote}.
Monomer Hamiltonian matrix elements were computed in
the GPU-accelerated \textsc{TeraChem} program
\cite{Ufimtzsev:2009:2619,Luehr:2011:949,Isborn:2012:5092} for classical
electronic structure theory, using TDA-TD-DFT \cite{Runge:1984:997} at
$\omega$PBE($\omega=0.3$)/6-31G* \cite{Tawada:2004:8425,Vydrov:2006:234109}.
Dimer Hamiltonian matrix elements were approximated by the
dipole/transition-dipole model. Dimer Hamiltonian matrix elements were truncated
after cyclical nearest-neighbor contacts due to the $r_{AB}^{-3}$ decay of the
interactions.  Figure \ref{fig:abs} depicts the simulated absorption spectrum of
this \emph{ab initio} exciton model computed from the excitation energies and
oscillator strengths of the lowest 18 electronic transitions with MC-VQE and
CIS, and compared to the ``full configuration interaction'' (FCI) reference
computed in the space of all possible $2^N$ monomer excitation configurations.
The CIS absorption shows a noticeable blue shift of a few hundredths of an eV
relative to FCI, and, more noticeably, the CIS oscillator strengths may deviate
by $10\%$ or more, particularly for the brightest states. By contrast, MC-VQE
with a single entangler layer is visually indistinguishable from FCI - the
maximum deviations of excitation energies are on the order of tens of $\mu$eV,
while the oscillator strengths generally deviate by $\ll1\%$. 
At the request of a reviewer, we have also considered a test case where CIS
produces qualitatively incorrect results relative to FCI: an $N=8$ linear stack
of BChl-a chromophores. MC-VQE has no trouble with this system, and again
produces results which are essentially visually indistinguishable from FCI: see
the Supplemental Material for full details \cite{SuppNote}.

\textit{Outlook } - In this Letter, we have demonstrated a hybrid
quantum/classical approach for the modeling of electronic absorption spectra in
large-scale photoactive complexes by using a multistate, contracted variant of
VQE (MC-VQE) in the context of an \emph{ab initio} exciton model.  We simulated
MC-VQE for an $N=18$ LH2 B850 complex (a Hilbert space
dimension of $2^{18}=262144$).
The MC-VQE
absorption spectrum matches FCI quantitatively with only a single layer of VQE
two-body entanglers with a connectivity matching that of the exciton
Hamiltonian. With a qubit count equivalent to the number of monomers $N$, a
circuit depth that is linear in $N$, a gate count that is quadratic in $N$, and
a requirement of only 1- and a sparse set of 2-body Pauli measurements, MC-VQE
applied to an \emph{ab initio} exciton model with local Hamiltonian connectivity is a
compelling application for deployment to near-term quantum hardware.

This Letter is intended to sketch the salient features of the MC-VQE algorithm
and its potential application to the \emph{ab initio} exciton model. Future work
will investigate robustness of the algorithm on realistic hardware including the
influence of gate/measurement errors. \emph{Ab initio} exciton Hamiltonians with
more-complicated local connectivity that are unlikely to be addressable with
classical methods such as DMRG should also be investigated.  Beyond this, effort
should be devoted to direct implementation on real hardware, where circuit
locality and simplification/sparsification will be of key importance. Finally,
MC-VQE should be explored in the context of direct simulation of fermionic
electronic structure problems - it seems highly likely that this algorithm will
be easily adaptable to the study of multiple excited states in many types of
Hamiltonians beyond the \emph{ab initio} exciton model.

\textbf{Note added during revision:} After the first revision of our paper was
released, we learned of the ``subspace search'' VQE (SS-VQE) approach developed
by Nakanishi, Mitarai, and Fujii in a recent preprint \cite{Nakanishi:2018:X}.
Both SS-VQE and MC-VQE use a state-averaged VQE entangler $\hat U$, and both
describe how to compute transition properties. The methods have several key
differences: SS-VQE uses hybrid quantum-classical optimization to determine the
minimal and maximal eigenvectors in the subspace matrix, while MC-VQE uses
classical diagonalization of the subspace Hamiltonian to determine all subspace
eigenstates simultaneously. Additionally, MC-VQE uses contracted reference
states (e.g., from CIS) while SS-VQE uses Hilbert-space configurations.

\textbf{Acknowledgements:} This material is based on work partially supported by
the U.S. Department of Energy, Office of Science, Office of Advanced Scientific
Computing Research, Scientific Discovery through Advanced Computing (SciDAC)
program.

\textbf{Financial Disclosure:} TJM is a cofounder of \textsc{PetaChem LLC}.
RMP and PLM own stock/options in \textsc{QC Ware Corp.}

\section{Supplemental Material}

\section{Hamiltonian Manipulation}

Consider an \emph{ab initio} exciton model of $N$ chromophoric monomers labelled
by index $A$, each with two monomer electronic states labeled $|0_A\rangle$
(ground) and $|1_A\rangle$ (excited). The \emph{ab initio} exciton Hamiltonian
is
\begin{equation}
\label{eq:H}
\hat H
=
\hat H^{(1)}
+
\hat H^{(2)}
\end{equation}
\[
=
\sum_{A}
\sum_{p,q \in 0,1}
(p_{A} | \hat h | q_{A})
| p_{A} \rangle
\langle q_{A} |
\]
\[
+
\sum_{A>B}
\sum_{p,q,r,s \in 0,1}
(p_{A} q_{A} | \hat v | r_{B} s_{B})
|p_{A} \rangle
\langle q_{A} |
\otimes
|r_{B} \rangle
\langle s_{B} |
\]
The real one-body matrix elements $(p_A|\hat h|q_A)$ are usually the 
(diagonal) adiabatic energy levels of the ground and first excited state of the
isolated monomers, while the real two-body matrix elements $(p_A q_A|\hat v|r_B
s_B)$ are the electrostatic interactions between the one-body densities (e.g.,
for $p = q$) or one-body transition densities (e.g., for $p \neq q$) of monomers
$A$ and $B$.  These matrix elements can be computed accurately with polynomial
cost (albeit expensive in absolute/prefactor considerations and requiring
extensive efforts to accelerate) by \emph{ab initio} electronic structure
computations performed on classical computers, for monomers with at least
several hundred atoms. 

\textbf{Pauli Operator Notation:} The one-body operator is,
\begin{equation}
\hat H^{(1)}
=
\sum_{A}
(0_A | \hat h | 0_A) | 0_A \rangle \langle 0_A |
+
(1_A | \hat h | 1_A) | 1_A \rangle \langle 1_A |
\end{equation}
\[
+
(0_A | \hat h | 1_A) | 0_A \rangle \langle 1_A |
+
(1_A | \hat h | 0_A) | 1_A \rangle \langle 0_A |
\]
\[
=
\sum_{A}
\frac{(0_A|\hat h|0_A) + (1_A|\hat h|1_A)}{2} \hat I_{A}
\]
\[
+
\frac{(0_A|\hat h|0_A) - (1_A|\hat h|1_A)}{2} \hat Z_{A}
\]
\[
+ (0_A|\hat h|1_A) \hat X_{A}
\]
\[
\equiv
\sum_{A}
S_A \hat I_{A}
+
D_A \hat Z_{A}
+
X_A \hat X_{A}
\]
Here $S_{A} = [(0_A | \hat h | 0_A) +
(1_A | \hat h | 1_A)] / 2$, $D_{A} \equiv [(0_A | \hat h | 0_A) - (1_A | \hat h
| 1_A)] / 2$, and $X_{A} = (0_A | \hat h | 1_A)$. When switching to Pauli matrix
notation on the second line, we use the implicit convention that any matrices
not shown are taken to be the identity, e.g., $\hat Z_{C} \equiv \hat I_{A}
\otimes \hat I_{B} \otimes \hat Z_{C} \otimes I_{D} \otimes \ldots \otimes \hat
I_{N}$. Note that the off-diagonal intramonomer couplings $X_A = (0_A | \hat h |
1_A)$ are usually defined to be zero in our exciton model convention, but there
is no overhead for including them here (there will be another $\hat X_A$ gate
contribution below).

The two-body operator $\hat H^{(2)}$ requires a bit more work to convert to a
form containing only sums of tensor products of Pauli operators. For each
two-body term,
\begin{equation}
H_{AB}
=
\end{equation}
\[
  (H_A | H_B) \hat H_A \otimes \hat H_B
+ (H_A | T_B) \hat H_A \otimes \hat X_B
\]
\[
+ (T_A | H_B) \hat X_A \otimes \hat H_B
+ (T_A | T_B) \hat X_A \otimes \hat X_B
\]
\[
+ (H_A | P_B) \hat H_A \otimes \hat P_B
+ (P_A | H_B) \hat P_A \otimes \hat H_B
\]
\[
+ (T_A | P_B) \hat X_A \otimes \hat P_B
+ (P_A | T_B) \hat P_A \otimes \hat X_B
\]
\[
+ (P_A | P_B) \hat P_A \otimes \hat P_B
\]
Here $H$, $P$, and $T$ represent ``hole,'' ``particle,'' and ``transition,''
respectively. $\hat H \equiv [\hat I + \hat Z] / 2$ and $\hat P \equiv [\hat I
- \hat Z] / 2$ are the hole and particle counting operators, respectively. Note
  that $\hat I = \hat H + \hat P$ and $\hat Z = \hat H - \hat P$ The two-body
matrix elements are,
\begin{equation}
(H_A | H_B) 
=
(0_A 0_A | \hat v | 0_B 0_B)
\end{equation}
\begin{equation}
(H_A | T_B) 
=
(0_A 0_A | \hat v | 0_B 1_B)
=
(0_A 0_A | \hat v | 1_B 0_B)
\end{equation}
\begin{equation}
(H_A | P_B)
=
(0_A 0_A | \hat v | 1_B 1_B)
\end{equation}
and so forth.

Terms 2, 3, 7, and 8 above can be written as,
\begin{equation}
+ (H_A | T_B) \hat H_A \otimes \hat X_B
+ (T_A | H_B) \hat X_A \otimes \hat H_B
\end{equation}
\[
+ (T_A | P_B) \hat X_A \otimes \hat P_B
+ (P_A | T_B) \hat P_A \otimes \hat X_B
\]
\[
=
  (S_A | T_B) \hat I_A \otimes X_B
+ (D_A | T_B) \hat Z_A \otimes X_B
\]
\[
=
+ (T_A | S_B) \hat X_A \otimes I_B
+ (T_A | D_B) \hat X_A \otimes Z_B
\]

Terms 1, 5, 6, and 9 above can be written as,
\begin{equation}
  (H_A | H_B) \hat H_A \otimes \hat H_B
+ (H_A | P_B) \hat H_A \otimes \hat P_B
\end{equation}
\[
+ (P_A | H_B) \hat P_A \otimes \hat H_B
+ (P_A | P_B) \hat P_A \otimes \hat P_B
\]
\[
=
  (S_A | S_B) \hat I_A \otimes \hat I_B
+ (S_A | D_B) \hat I_A \otimes \hat Z_B
\]
\[
+ (D_A | S_B) \hat Z_A \otimes \hat I_B
+ (D_A | D_A) \hat Z_A \otimes \hat Z_B
\]

In the above,
\begin{equation}
(S_A | T_B) = (H_A + P_A | T_B) / 2 
\end{equation}
\[
= [(H_A | T_B) + (P_A | T_B)] / 2
\]
\begin{equation}
(D_A | T_B) = (H_A - P_A | T_B) / 2 
\end{equation}
\[
= (H_A | T_B) - [(P_A | T_B)] / 2
\]
and so forth.

So the two-body Hamiltonian element can be written as,
\begin{equation}
\hat H_{AB}^{(2)} = 
   (S_A | S_B) \hat I_A \otimes \hat I_B
\end{equation}
\[
+  (S_A | D_B) \hat I_A \otimes \hat Z_B
+  (D_A | S_A) \hat Z_A \otimes \hat I_B
\]
\[
+  (S_A | T_B) \hat I_A \otimes \hat X_B
+  (T_A | S_A) \hat X_A \otimes \hat I_B
\]
\[
+  (T_A | T_B) \hat X_A \otimes \hat X_B
+  (T_A | D_B) \hat X_A \otimes \hat Z_B
\]
\[
+  (D_A | T_B) \hat Z_A \otimes \hat X_B
+  (D_A | D_B) \hat Z_A \otimes \hat Z_B
\]

\textbf{Hamiltonian in Pauli Notation:} After the straightforward
algebra above, the total Hamiltonian can be written as,
\begin{equation}
\label{eq:H_N}
\hat H
=
\mathcal{E}
+
\mathcal{H}^{(1)}
+
\mathcal{H}^{(2)}
=
\mathcal{E} \hat I 
+ 
\sum_{A}
\mathcal{Z}_{A}
\hat Z_{A}
+
\mathcal{X}_{A}
\hat X_{A}
\end{equation}
\[
+
\sum_{A > B}
\mathcal{XX}_{AB}
\hat X_{A} \otimes \hat X_{B}
+
\mathcal{XZ}_{AB}
\hat X_{A} \otimes \hat Z_{B}
\]
\[
+
\mathcal{ZX}_{AB}
\hat Z_{A} \otimes \hat X_{B}
+
\mathcal{ZZ}_{AB}
\hat Z_{A} \otimes \hat Z_{B}
\]
The matrix elements are,
\begin{equation}
\mathcal{E}
=
\sum_{A}
S_{A}
+
\sum_{A>B}
(S_A | S_B)
\end{equation}
\begin{equation}
\mathcal{Z}_{A}
=
D_{A}
+
\sum_{B}
(D_{A} | S_{B})
\end{equation}
\begin{equation}
\mathcal{X}_{A} 
=
X_{A}
+
\sum_{B}
(T_{A} | S_{B})
\end{equation}
\begin{equation}
\mathcal{XX}_{AB}
=
(T_{A} | T_{B})
\end{equation}
\begin{equation}
\mathcal{XZ}_{AB}
=
(T_{A} | D_{B})
\end{equation}
\begin{equation}
\mathcal{ZX}_{AB}
=
(D_{A} | T_{B})
\end{equation}
\begin{equation}
\mathcal{ZZ}_{AB}
=
(D_{A} | D_{B})
\end{equation}
More specifically,
\begin{equation}
S_{A} = 
\left [
(0_A | \hat h |0_A)
+
(1_A | \hat h |1_A)
\right ] / 2
\end{equation}
\begin{equation}
D_{A} = 
\left [
(0_A | \hat h |0_A)
-
(1_A | \hat h |1_A)
\right ] / 2
\end{equation}
\begin{equation}
X_{A} =
(0_A | \hat h | 1_A)
\end{equation}
\begin{equation}
(T_A | T_B) 
=
(0_A 1_A | \hat v | 0_B 1_B)
\end{equation}
\begin{equation}
(T_A | S_B) 
=
(0_A 1_A | \hat v | 0_B 0_B + 1_B 1_B) / 2
\end{equation}
\begin{equation}
(T_A | D_B) 
=
(0_A 1_A | \hat v | 0_B 0_B - 1_B 1_B) / 2
\end{equation}
\begin{equation}
(S_A | T_B) 
=
(0_A 0_A + 1_A 1_A | \hat v | 0_B 1_B) / 2
\end{equation}
\begin{equation}
(D_A | T_B) 
=
(0_A 0_A - 1_A 1_A | \hat v | 0_B 1_B) / 2
\end{equation}
\begin{equation}
(S_A | S_B)
=
(0_A 0_A + 1_A 1_A | \hat v | 0_B 0_B + 1_B 1_B) / 4
\end{equation}
\begin{equation}
(S_A | D_B)
=
(0_A 0_A + 1_A 1_A | \hat v | 0_B 0_B - 1_B 1_B) / 4
\end{equation}
\begin{equation}
(D_A | S_B)
=
(0_A 0_A - 1_A 1_A | \hat v | 0_A 0_B + 1_A 1_B) / 4
\end{equation}
\begin{equation}
(D_A | D_B)
=
(0_A 0_A - 1_A 1_A | \hat v | 0_A 0_B - 1_A 1_B) / 4
\end{equation}
Note that $\{ \mathcal{Z}_{A} \}$ are expected to be the largest matrix elements
in this Hamiltonian, and are of the order of a few eV ($10^{-2} - 10^{-1}$ au).
In practice, we typically find the $\{ \mathcal{X}_{A} \}$ and $\{
\mathcal{XX}_{AB} \}$ are the next largest matrix elements ($10^{-3} - 10^{-2}$
au). These are rough estimates which may differ in specific systems.

\section{MC-VQE Technical Details}

\textbf{CIS State Preparation:} An important task for excited state MC-VQE is to
prepare the configuration interaction singles (CIS) state,
\[
| \Phi_{\Theta} \rangle
\equiv
\mu | 000 \ldots \rangle
+
\alpha | 100\ldots \rangle
+
\beta | 010\ldots \rangle
+
\gamma | 001\ldots \rangle
+
\ldots
\]
\begin{equation}
:
\sqrt{
\mu^2
+
\alpha^2
+
\beta^2
+
\gamma^2
\ldots
}
=
1
\end{equation}
The parameters $\mu, \alpha, \beta, \gamma \ldots$ are obtained classically by
solving the CIS eigenproblem. Note that the CIS eigenstates are definitionally
orthonormal,
\begin{equation}
\langle \Phi_{\Theta} | \Phi_{\Theta'} \rangle
=
\delta_{\Theta, \Theta'}
, \
\Theta, \Theta' \in [0, N]
\end{equation}
Note that there are $N+1$ states in the CIS manifold used in this work: $1$ for
the reference ground configuration ($|000 \ldots \rangle$) and $N$ for the
singly-excited configurations ($|100 \ldots \rangle$ and so forth), all of which
are allowed to mix in our CIS prescription.

A circuit to prepare the CIS state is sketched for $N=5$,
\begin{widetext}
\begin{equation}
\Qcircuit @R=1.0em @C=1.0em {
\lstick{|0_A\rangle}
 & \gate{R_{y} (\theta_0)}
 & \ctrl{1}
 & \qw
 & \qw
 & \qw
 & \qw
 & \qw
 & \qw
 & \qw
 & \qw
 & \qw
 & \targ
 & \targ
 & \targ
 & \targ
 & \qw \\
\lstick{|0_B\rangle}
 & \qw
 & \gate{F_{y} (\theta_{AB}) }
 & \ctrl{1}
 & \qw
 & \qw
 & \qw
 & \qw
 & \qw
 & \targ
 & \targ
 & \targ
 & \qw
 & \qw
 & \qw
 & \ctrl{-1}
 & \qw \\
\lstick{|0_C\rangle}
 & \qw
 & \qw
 & \gate{F_{y} (\theta_{BC}) }
 & \ctrl{1}
 & \qw
 & \qw
 & \targ
 & \targ
 & \qw
 & \qw
 & \ctrl{-1}
 & \qw
 & \qw
 & \ctrl{-2}
 & \qw
 & \qw \\
\lstick{|0_D\rangle}
 & \qw
 & \qw
 & \qw
 & \gate{F_{y} (\theta_{CD}) }
 & \ctrl{1}
 & \targ
 & \qw
 & \ctrl{-1}
 & \qw
 & \ctrl{-2}
 & \qw
 & \qw
 & \ctrl{-3}
 & \qw
 & \qw
 & \qw \\
\lstick{|0_E\rangle}
 & \qw
 & \qw
 & \qw
 & \qw
 & \gate{F_{y} (\theta_{DE}) }
 & \ctrl{-1}
 & \ctrl{-2}
 & \qw
 & \ctrl{-3}
 & \qw
 & \qw
 & \ctrl{-4}
 & \qw
 & \qw
 & \qw
 & \qw 
\\
}
\end{equation}
\end{widetext}
The $R_y (\Theta_0)$ ``pump'' gate always goes with $|0_A\rangle$, otherwise the
circuit nests ``matryoshka'' style.

The controlled $F_y (\theta)$ gate is,
\begin{widetext}
\begin{equation}
\Qcircuit @R=1.0em @C=1.0em {
& \ctrl{1} & \qw  \\
& \gate{F_{y} (\theta)} & \qw \\
}
=
\Qcircuit @R=1.0em @C=1.0em {
& \qw & \ctrl{1} & \qw & \qw  \\
& \gate{R_{y} (-\theta/2)} & \gate{Z} & \gate{R_{y} (+\theta/2)} & \qw \\
}
=
\Qcircuit @R=1.0em @C=1.0em {
& \qw & \qw & \ctrl{1} & \qw & \qw & \qw  \\
& \gate{R_{y} (-\theta/2)} & \gate{H} & \targ & \gate{H} & \gate{R_{y} (+\theta/2)} & \qw \\
}
=
\left [
\begin{array}{cccc}
 1 & & & \\
 & 1 & & \\
 & & +c(\theta) & +s(\theta) \\
 & & +s(\theta) & -c(\theta) \\
\end{array}
\right ]
\end{equation}
\end{widetext}

By inspection,
\begin{equation}
\mu
=
\cos \theta_{0}
\end{equation}
\begin{equation}
\alpha
=
\sin \theta_{0}
\cos \theta_{AB}
\end{equation}
\begin{equation}
\beta
=
\sin \theta_{0}
\sin \theta_{AB}
\cos \theta_{BC}
\end{equation}
\begin{equation}
\gamma
=
\sin \theta_{0}
\sin \theta_{AB}
\sin \theta_{BC}
\cos \theta_{CD}
\end{equation}
\begin{equation}
\vdots
\end{equation}
\begin{equation}
\eta
=
\sin \theta_{0}
\ldots
\sin \theta_{LM}
\sin \theta_{MN}
\end{equation}
Thus we can evaluate the angles recursively,
\begin{equation}
\theta_{0}
=
\cos^{-1} (\mu)
\end{equation}
\begin{equation}
\theta_{AB}
=
\cos^{-1} (\alpha / (\sin(\theta_{0})))
\end{equation}
\begin{equation}
\theta_{BC}
=
\cos^{-1} (\beta / (\sin(\theta_{0}) \sin(\theta_{AB})))
\end{equation}
\begin{equation}
\theta_{CD}
=
\cos^{-1} (\gamma / (\sin(\theta_{0}) \sin(\theta_{AB}) \sin(\theta_{BC})))
\end{equation}
\begin{equation}
\vdots
\end{equation}
\begin{equation}
\theta_{MN}
=
\cos^{-1} (\mu / (\sin(\theta_{0}) \sin(\theta_{AB}) \sin(\theta_{BC}) \ldots \sin(\theta_{LM})))
\end{equation}
Note that the last angle $\theta_{MN}$ is only resolved up to a phase factor of
$\pm 1$: the explicit value of $\eta$ must be checked using the last line of the
forward formula above to determine the phase factor (the relative error will be
0 [+1] or 2 [-1], which makes the check particularly easy).

The ``interference'' state $(|\Phi_{\Theta}\rangle \pm | \Phi_{\Theta'} \rangle)
/ \sqrt{2}$ can easily be formed by substitutions of the coefficients
\begin{equation}
\mu^{\pm}
=
(
\mu^{\Theta}
\pm
\mu^{\Theta'}
) / \sqrt{2}
\end{equation}
\begin{equation}
\alpha^{\pm}
=
(
\alpha^{\Theta}
\pm
\alpha^{\Theta'}
) / \sqrt{2}
\end{equation}
\begin{equation}
\vdots
\end{equation}

Note that the CIS state with $\mu = 0, \alpha = \beta = \gamma = \ldots = 1 /
\sqrt{N}$ is usually called the $|W_N\rangle$ state. A circuit for the $|W_N
\rangle$ that was used as the basis for this work is presented in
\cite{Diker:2016:W}.

\textbf{VQE Global Entangler Matrix:} In any configuration-space basis, the
adiabatic eigenfunctions of the real electronic or \emph{ab initio} exciton
Hamiltonian can be written as real, orthonormal vectors with arbitrary total
phase of $\pm 1$. Therefore, the VQE entangler operator $\hat U$ can be
restricted to $SO(2^N)$ without loss.  We have elected to construct the total VQE
entangler circuit for the \emph{ab initio} exciton model by placing a two-body
entangler restricted to $SO(4)$ at each two-body interaction site in the exciton
Hamiltonian. E.g., for a linear arrangement,
\begin{widetext}
\begin{equation}
\label{eq:vqe-top}
\Qcircuit @R=1.0em @C=1.0em {
\lstick{| A \rangle}
& \multigate{1}{U_{2}^{AB}}
& \qw
& \qw \\
\lstick{| B \rangle}
& \ghost{U_{2}^{AB}}
& \multigate{1}{U_{2}^{BN}}
& \qw \\
\lstick{| N \rangle}
& \multigate{1}{U_{2}^{NZ}}
& \ghost{U_{2}^{BN}}
& \qw \\
\lstick{| Z \rangle}
& \ghost{U_{2}^{NZ}}
& \qw
& \qw \\
}
\end{equation}
\end{widetext}
If additional variational flexibility in the ansatz is desired, a
straightforward approach is to add additional layers of entanglers of the form
shown here, or to extend two-body entanglers to the next layer(s) of nearest
neighbors.

Details of the specific VQE two-body entangler $\hat U_2$ restricted to $SO(4)$
that is used in this work are presented below - these manipulations are intended
to produce a VQE two-body entangler whose parameters are easy to guess (e.g.,
starting from all zeros) and to optimize with partial gradient information
(e.g., by simple gradient-descent or L-BFGS) for problems encountered in the
\emph{ab initio} exciton model.

Note that there has been much interest in the literature on the construction of
optimal 2-body quantum circuits covering $SU(4)$ or $SO(4)$ using various or
arbitrary gate libraries - for an overview, see
\cite{Zhang:2003:027903,Shende:2004:062321,Vatan:2004:032315,Wei:2012:X}. 

\textbf{VQE Two-Body Entangler Matrix:} $SO(4)$ is the group of real, orthogonal
matrices with determinant $+1$, and covers all possible two-body entangler
matrices needed in our VQE task. There are infinitely many equivalent logical
parametrizations of $SO(4)$ with 6 real parameters, but some of these will prove
easier to optimize than others in VQE applications. 

For instance, one particularly straightforward parametrization of $SO(4)$ is
given by,
\begin{equation}
\label{eq:so4mark1}
\hat U_2
=
\exp
\left (
\left [
\begin{array}{cccc}
 0 & +A & +B & +C \\
-A &  0 & +D & +E \\
-B & -D &  0 & +F \\
-C & -E & -F &  0 \\
\end{array}
\right ]
\right )
\end{equation}
That is, any special orthogonal matrix for $N=4$ can be written as the matrix
exponential of an $N=4$ antisymmetric matrix with 6 unconstrained real
parameters $A$ through $F$.

Another, equivalent parametrization can be realized by considering the two-body
Pauli generators of the antisymmetric group,
\begin{equation}
-i \hat Y_{A} \otimes \hat I_{B}
=
\left [
\begin{array}{cccc}
 0 &  0 & -1 &  0 \\
 0 &  0 &  0 & -1 \\
+1 &  0 &  0 &  0 \\
 0 & +1 &  0 &  0 \\
\end{array}
\right ]
\end{equation}
\begin{equation}
-i \hat Y_{A} \otimes \hat X_{B}
=
\left [
\begin{array}{cccc}
 0 &  0 &  0 & -1 \\
 0 &  0 & -1 &  0 \\
 0 & +1 &  0 &  0 \\
+1 &  0 &  0 &  0 \\
\end{array}
\right ]
\end{equation}
\begin{equation}
-i \hat Y_{A} \otimes \hat Z_{B}
=
\left [
\begin{array}{cccc}
 0 &  0 & -1 &  0 \\
 0 &  0 &  0 & +1 \\
+1 &  0 &  0 &  0 \\
 0 & -1 &  0 &  0 \\
\end{array}
\right ]
\end{equation}
\begin{equation}
-i \hat I_{A} \otimes \hat Y_{B}
=
\left [
\begin{array}{cccc}
 0 & -1 &  0 &  0 \\
+1 &  0 &  0 &  0 \\
 0 &  0 &  0 & -1 \\
 0 &  0 & +1 &  0 \\
\end{array}
\right ]
\end{equation}
\begin{equation}
-i \hat X_{A} \otimes \hat Y_{B}
=
\left [
\begin{array}{cccc}
 0 &  0 &  0 & -1 \\
 0 &  0 & +1 &  0 \\
 0 & -1 &  0 &  0 \\
+1 &  0 &  0 &  0 \\
\end{array}
\right ]
\end{equation}
\begin{equation}
-i \hat Z_{A} \otimes \hat Y_{B}
=
\left [
\begin{array}{cccc}
 0 & -1 &  0 &  0 \\
+1 &  0 &  0 &  0 \\
 0 &  0 &  0 & +1 \\
 0 &  0 & -1 &  0 \\
\end{array}
\right ]
\end{equation}
Therefore, 
\begin{equation}
\label{eq:so4mark2}
\hat U_2
=
\exp(
\end{equation}
\[
-i \theta_{XY} \hat X_{A} \otimes \hat Y_{B}
-i \theta_{ZY} \hat Z_{A} \otimes \hat Y_{B}
-i \theta_{YZ} \hat Y_{A} \otimes \hat Z_{B}
\]
\[
-i \theta_{YX} \hat Y_{A} \otimes \hat X_{B}
-i \theta_{YI} \hat Y_{A} \otimes \hat I_{B}
-i \theta_{IY} \hat I_{A} \otimes \hat Y_{B}
)
\]
with the six real parameters
$\theta_{IY}$,
$\theta_{YI}$,
$\theta_{XY}$,
$\theta_{YX}$,
$\theta_{ZY}$, and
$\theta_{YZ}$.
The correspondence between parameter sets is,
\begin{equation}
A = -(\theta_{IY} + \theta_{ZY})
\end{equation}
\begin{equation}
F = -(\theta_{IY} - \theta_{ZY})
\end{equation}
\begin{equation}
C = -(\theta_{YX} + \theta_{XY})
\end{equation}
\begin{equation}
D = -(\theta_{YX} - \theta_{XY})
\end{equation}
\begin{equation}
B = -(\theta_{YI} + \theta_{YZ})
\end{equation}
\begin{equation}
E = -(\theta_{YI} - \theta_{YZ}) 
\end{equation}
We have found that the second parametrization of the VQE two-body entangler is
sometimes easier to optimize than the first, e.g., when using straightforward
gradient descent with finite-difference gradients, and using an initial guess of
zero entanglement (all angles $\{ \theta_{IY} \ldots \theta_{YZ} \}$ or
antisymmetric generator parameters $\{ A \ldots F \}$ set to zero). With tightly
converged optimizations using L-BFGS, both logical parametrizations of the
two-body entanglers produce results with similar accuracy in \emph{ab initio}
exciton model excitation energies and oscillators strengths.

We note that there are many possible physical realizations of two-body
entangler gates covering $SO(4)$, depending on the specific gate library of a
given quantum computer. For instance, one possible two-body entangler circuit
is \cite{Wei:2012:X},
\begin{equation}
\label{eq:so4mark4}
\Qcircuit @R=1.0em @C=0.5em {
\lstick{|0_A\rangle}
 & \gate{R_{y} (\theta_{1})}
 & \ctrl{1}
 & \gate{R_{y} (\theta_{3})}
 & \ctrl{1}
 & \gate{R_{y} (\theta_{5})}
 & \qw \\
\lstick{|0_B\rangle}
 & \gate{R_{y} (\theta_{2})}
 & \targ
 & \gate{R_{y} (\theta_{4})}
 & \targ
 & \gate{R_{y} (\theta_{6})}
 & \qw \\
}
\end{equation}
In practice, it seems important to consider different possible logical
parametrizations of the two-body entangler circuits, regardless of the
underlying physical parametrization. To highlight this, we note that direct
optimization of the two-body entanglers in terms of the angles $\{ \theta_{1}
\ldots \theta_{6} \}$ in Equation \ref{eq:so4mark4} did not converge to the same
qualitative solution as Equations \ref{eq:so4mark1} or \ref{eq:so4mark2} above,
when using steepest descent (though this circuit converged to a local minimum
without apparent convergence difficulties). In fact, the solution obtained with
the parametrization of Equation \ref{eq:so4mark4} was not qualitatively superior
to the CIS polynomial-scaling classical solution. Since the circuit in Equation
\ref{eq:so4mark4} covers $SO(4)$ just as well as the logical parametrizations of
Equations \ref{eq:so4mark1} or \ref{eq:so4mark2}, it likely the case that the
all-zeros guess lies in the neighborhood of an undesired local minimum.
Switching to L-BFGS, the two-body entanglers of Equation \ref{eq:so4mark4}
eventually converge to the same qualitatively solution as in Equations
\ref{eq:so4mark1} or \ref{eq:so4mark2}, but some convergence difficulties are
encountered, and a the L-BFGS optimization takes approximately twice as many
iterations. The selection of an optimal logical parametrization of the entangler
circuits, mapping to physical parametrization in terms of available gate
library, and design of algorithms for robust convergence to the desired MC-VQE
minimum is obviously a subject worth further consideration in future work.

All results depicted in the primary manuscript use a logical two-body entangler
circuit parametrized by Equation \ref{eq:so4mark2}, and optimized by L-BFGS with
finite-difference gradients (second-order symmetric with $\Delta \theta =0.01$),
and using an initial guess of zero entanglement with all angles $\{ \theta_{IY}
\ldots \theta_{YZ} \}$ starting at zero.

Example output files of the optimization profiles with steepest descent and
L-BFGS and using the various parametrizations of the two-body entanglers shown
above are provided for an $N=12$ subset of the B850 ring complex in the
supplementary data packet.

%

\section{Computational Details}

All computational results for the B850 ring were run with \textsc{Quasar} GIT
SHA 4ce64a05a0f5f63988cb68209d24e085d2de275f.

The XYZ geometry of the $N=18$ BChl-$a$ LH2 B850 ring complex is contained in
the supplemental data packet. Monomer energies, densities, and
dipole/transition-dipole moments (relative to center of mass) are computed for
each monomer in isolation using $\omega$PBE($\omega=0.3$)/6-31G*. Ground-state
DFT is used to compute the ground state, Tamm-Dancoff Approximation TD-DFT
(TDA-TD-DFT) is used to compute the excited state. The $S_1$ total and $S_0-S_1$
transition dipole moment are computed without orbital response (unrelaxed).
Two-body interaction matrix elements are computed between monomers using the
dipole-dipole interaction formula,
\begin{equation}
V_{AB}
=
\frac{
\vec \mu_{A} \cdot \vec \mu_{B}
-
3
(\vec \mu_A{} \cdot \vec n_{AB})
(\vec \mu_B{} \cdot \vec n_{AB})
}{r_{AB}^3}
\end{equation}
Here $\vec \mu_{A}$ is the total or transition dipole moment, $\vec n_{AB}$ is
the normal vector along the displacement between the centers of mass of monomers
$A$ and $B$, and $r_{AB}$ is the distance between the centers of mass of
monomers $A$ and $B$.

The dipole oscillator strength is computed from the energy gaps and transition
dipole moments between approximate eigenstates,
\begin{equation}
O_{\Theta \Theta'}
=
\frac{2}{3}
(E_{\Theta'} - E_{\Theta})
\langle
\Psi_{\Theta} |
\hat \mu
| \Psi_{\Theta'}
\rangle^2
\end{equation}
The dipole operators can be written as sums of 1-body Pauli operators, e.g.,
\begin{equation}
\hat \mu
\equiv
\sum_{A}
\vec \mu_{\hat I}^A
\hat I_{A}
+
\vec \mu_{\hat Z}^A
\hat Z_{A}
+
\vec \mu_{\hat X}^A
\hat X_{A}
\end{equation}
where $\vec \mu_{I}^A = (\vec \mu_{A}^{11} + \vec \mu_{A}^{00}) / 2$, $\vec
\mu_{Z}^A = (\vec \mu_{A}^{11}
- \vec \mu_{A}^{00}) / 2$, and $\vec \mu_{X}^{A} = \vec \mu_{A}^{01}$ are
  computed from the monomer total and transition dipole moments.

To aid in replication of the results the total energies of the $S_0$ and $S_1$
states of each monomer are provided in Table \ref{tab:E}. The classical monomer
quantum chemistry outputs from \textsc{TeraChem} and full numerical details of
the \emph{ab initio} exciton Hamiltonian for this system are present in the
supplemental data packet. For the full $N=18$ BChl-$a$ B850 ring complex system
in LH2, the L-BFGS iterative history, optimized MC-VQE parameter values, and
other characteristics of the converged MC-VQE solution are present in the
\textsc{Quasar} output file and \textsc{.NPZ} data file in the supplemental data
packet.

\begin{table}[h!]
\begin{center}
\caption{
Monomer $S_0$ and $S_1$ state energies for $N=18$ BChl-$a$ LH2 B850 ring
complex computed at $\omega$PBE($\omega=0.3$)/6-31G* using DFT and TDA-TD-DFT.}
\label{tab:E}
\begin{tabular}{lrr}
\hline \hline
$A$ & $E_{S_0}$ & $E_{S_1}$ \\
\hline
 1 & -2263.26377175 & -2263.19429617 \\
 2 & -2263.25985281 & -2263.18713945 \\
 3 & -2263.26107440 & -2263.18942087 \\
 4 & -2263.27390235 & -2263.20150610 \\
 5 & -2263.24311586 & -2263.17416282 \\
 6 & -2263.29019565 & -2263.21759301 \\
 7 & -2263.25812333 & -2263.18493523 \\
 8 & -2263.31462943 & -2263.24431121 \\
 9 & -2263.28358218 & -2263.21345739 \\
10 & -2263.23436184 & -2263.16413403 \\
11 & -2263.27574682 & -2263.20553908 \\
12 & -2263.28354903 & -2263.21341881 \\
13 & -2263.28916608 & -2263.21882588 \\
14 & -2263.27286521 & -2263.20307447 \\
15 & -2263.27213702 & -2263.20127737 \\
16 & -2263.29074598 & -2263.22183660 \\
17 & -2263.27479708 & -2263.20232022 \\
18 & -2263.29686920 & -2263.22028791 \\
\hline \hline
\end{tabular}
\end{center}
\end{table}

\section{Additional Case Study: BChl-a $H$-Aggregate Stack}

At the request of a reviewer, we have additionally considered a test case where
CIS performs markedly poorly, to verify that MC-VQE continues to provide
accurate results for systems with qualitatively more multi-excitonic character
than the simple $J$-aggregate-type B850 ring complex studied in the main
manuscript. To that end, we have considered a linear stack of $N=8$ truncated
BChl-a units, stacked in an aligned geometry and then allowed to geometrically
relax with $\omega$PBE($\omega=0.3$)/6-31G*-D3. An \emph{ab initio} exciton
model was constructed for this system using the same TDA-TD-DFT
$\omega$PBE($\omega=0.3$)/6-31G* treatment of the monomers and unrelaxed
dipole-dipole couplings as described elsewhere for the B850 ring complex. The
lowest 9 states and 8 ground-to-excited oscillator strengths were computed with
FCI, CIS, and MC-VQE. For the MC-VQE computations, a slightly updated version of
our \textsc{Quasar} code was used, with the same overall topology of two-layer
linear VQE entangler circuit depicted in Equation \ref{eq:vqe-top} and with the
two-body entanglers constructed as in \ref{eq:so4mark4}, and with redundant
$\hat R_y$ gates at the interface between the two layers of two-body entanglers
removed. The gradient-free Powell method was used to tightly optimize the
state-averaged VQE entangler circuit parameters to a maximum state-averaged
energy gradient of $1\times10^{-7}$. The specific geometry, \textsc{.NPZ} file
containing \emph{ab initio} exciton model matrix elements, \textsc{Quasar}
output file (including state-averaged VQE entangler circuit construction and
optimized parameters), and \text{.NPZ} file containing the output state energies
and oscillator strengths are provided in the supplemental data packet.

The stacking geometry of the chromophores in this case study promotes
$H$-aggregate-type electronic behavior, wherein the monomer states split and
generally blue shift, and the oscillator strengths of the lowest few states are
heavily attenuated relative to higher-lying states. Double and higher-lying
excitations strongly modulate the energies, characters, and oscillator strengths
of the adiabatic electronic states for this system. As such, CIS performs
extremely poorly relative to FCI for this system, as depicted in Figure 3: the
CIS excitation energies are all blue-shifted by between 0.1 and 0.5 eV from the
FCI excitation energies, and the CIS oscillator strengths are often
qualitatively incorrect (e.g., relative errors in the vicinity of 100\% for
states with significant oscillator strengths). Even the apparent agreement of
CIS and FCI oscillator strengths for the brightest transition between $S_0$ and
$S_7$ is accidental: examination of the $S_7$ excited state populations of the
monomers $P_{A}^{\Theta} \equiv \langle \Psi_{\Theta} | 1_A \rangle \langle 1_A
| \Psi_{\Theta} \rangle$ indicates that the majority of the exciton lives on
monomers 3 and 4 in zero-based ordering (with more-minor contributions on
monomers 0 and 1) in FCI, but the corresponding CIS state has the majority of
the exciton on monomer 2 and then spread to monomers 3, 4, and 5, i.e., the
state characters are markedly different. In more numerical terms, the fidelity
between FCI and CIS is $| \langle \Psi_{7}^{\mathrm{FCI}} |
\Psi_{7}^{\mathrm{CIS}} \rangle | = 0.688$, while the corresponding fidelity
between FCI and MC-VQE is $| \langle \Psi_{7}^{\mathrm{FCI}} |
\Psi_{7}^{\mathrm{VQE}} \rangle | = 0.995$.  Therefore, the agreement of
oscillator strengths between CIS and FCI for this state is purely accidental,
and CIS produces an absorption spectrum profile which is both qualitatively and
quantitatively incorrect. By contrast, MC-VQE (starting from the \emph{same} CIS
contracted reference states) produces and absorption spectrum profile which is
generally visually indistinguishable from FCI, and which agrees in the details
of the state characters, without requiring modifications to the topology of the
state-averaged VQE entangler circuit. The MC-VQE excitation energies are
generally in error with FCI by at most 0.01 eV, while the corresponding
oscillator strengths are generally in error with FCI by at most 0.1 [-].
Inspection of the excited state populations indicates that the state characters
are highly congruent between FCI and MC-VQE. The agreement of MC-VQE and FCI is
not quite as perfect as the $N=18$ B850 ring in the main text, but the
improvement of MC-VQE over CIS is much more striking.  Overall, this case study
provides compelling evidence that MC-VQE can provide reliable results for
difficult problems involving significant multi-excitonic character, all while
starting from poor-quality CIS contracted reference states and without modifying
the nature of the state-averaged VQE entangler circuit.

\begin{figure}
\begin{center}
\includegraphics[width=3.2in]{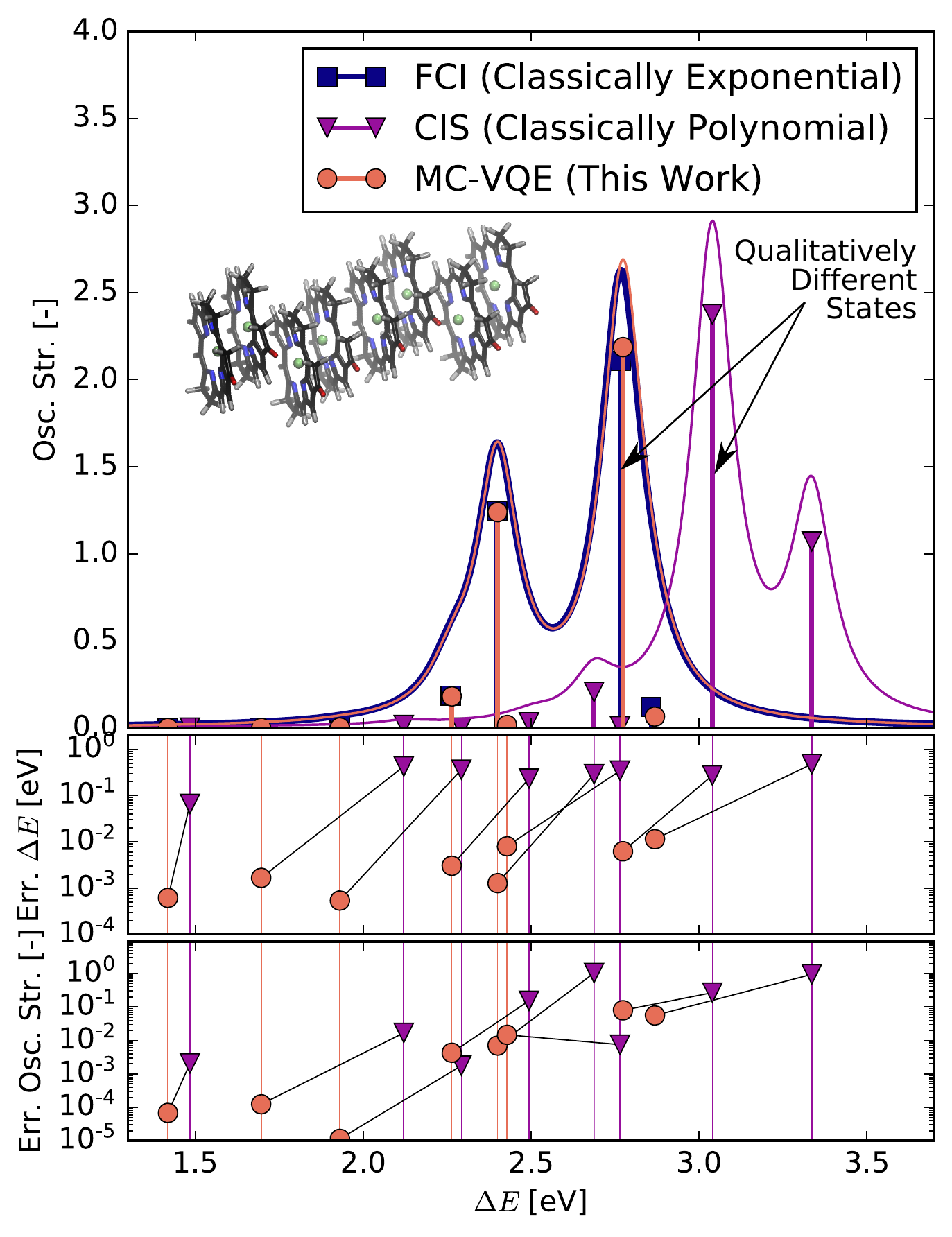}
\caption{(color online). Top - Simulated absorption spectrum of $N=8$ linear
stack BChl-a test case (geometry depicted in inset), computed from the excitation
energies and oscillator strengths of the lowest 8 electronic transitions,
depicted as vertical sticks.  The envelope of the absorption spectrum is
sketched by broadening the contribution from each transition with a Lorentzian
with width of $\delta=0.15$ eV. The simulated MC-VQE and reference FCI results
are largely visually indistinguishable. Middle
- errors in excitation energies. Bottom - errors in oscillator strengths. Middle
  and bottom - thin lines are a guide for the eye.
}
\label{fig:stack}
\end{center}
\end{figure}



%


\end{document}